\documentclass[3p,times,twocolumn]{elsarticle}

\usepackage{ecrc}

\volume{00}

\firstpage{1}
 \biboptions{comma,sort&compress}

\journalname{Nuclear and Particle Physics Proceedings}

\runauth{}

\jid{nppp}

\jnltitlelogo{Nuclear and Particle Physics Proceedings}

\usepackage{amsmath,amsfonts,amssymb}
\usepackage{hyperref}
\usepackage{epsfig}

\usepackage[figuresright]{rotating}

\def\bra#1{\langle #1 |\, }
\def\ket#1{\, | #1 \rangle}
\newcommand{\ba}{\begin{array}}
\newcommand{\ea}{\end{array}}

\newcommand{\be}{\begin{equation}}
\newcommand{\ee}{\end{equation}}
\newcommand{\beqn}{\begin{eqnarray}}
\newcommand{\eeqn}{\end{eqnarray}}

\newcommand{\ImPi}{\operatorname{Im}\Pi}

\newcommand{\oo}{\langle\mathcal{O}_{1}\rangle_{\mu}}
\newcommand{\oei}{\langle\mathcal{O}_{8}\rangle_{\mu}}

\begin{document}

\begin{frontmatter}

\dochead{}

\title{Confronting hadronic tau decays with non-leptonic kaon decays$^*$}
\cortext[cor0]{Talk given at 21th International Conference in Quantum Chromodynamics (QCD 18),  2 - 6 July 2018, Montpellier - FR}

\cortext[cor1]{Speaker.}
\author[label1,label2]{Antonio Rodr\' iguez-S\' anchez\corref{cor1}}
\author[label1]{Antonio Pich}
\address[label1]{Departament de F\'\i sica Te\`orica, IFIC, Universitat de Val\`encia -- CSIC,\\
Apt. Correus 22085, E-46071 Val\`encia, Spain}

\address[label2]{Department of Astronomy and Theoretical Physics,
Lund University,\\ S\" olvegatan 14A, SE 223-62 Lund, Sweden}

\begin{abstract}
In the chiral limit, the $D=6$ contribution to the Operator Product Expansion (OPE) of the $\mathrm{VV-AA}$ correlator of quark currents only depends on two vacuum condensates, which can be related to hadronic matrix elements associated to CP violation in non-leptonic kaon decays. We use those relations to determine $\langle(\pi\pi)_{I=2}|\mathcal{Q}_{8}|K\rangle$, using the updated ALEPH spectral functions. Alternatively, we use those relations in the opposite direction. Taking the values of the matrix elements from the lattice to obtain the $D=6$ vacuum elements provides a new short-distance constraint which allows for an inclusive determination of $f_{\pi}$ and an updated value for the $D=8$ condensate.
\end{abstract}


\end{frontmatter}

\section{Introduction}
\label{}
Hadronic tau decays are a gold mine to study and test many of the properties of the different interactions at their most fundamental (known) level \cite{Pich:2013lsa}. The weak nature of these decays, which occur through the interaction of a lepton and a quark charged current mediated by an off-shell $W$ (see Fig. \ref{fig:diagram}.), makes them a very interesting electroweak laboratory for some theoretically clean observables \cite{Cirigliano:2018dyk}. Additionally, the quarks generated in the decay hadronize. As a consequence of the additional neutrino emission, one observes a very rich hadronic continuum from energies where low-energy QCD methods such a $\chi \mathrm{PT}$ are valid, to the tau mass, where nearly perturbative methods can be applied, conforming a very nice window to strong interactions at different scales.

\begin{center}
\vspace*{-0.25cm}
{\begin{figure}[hbt]
\begin{center}
{\includegraphics[width=5.5cm]{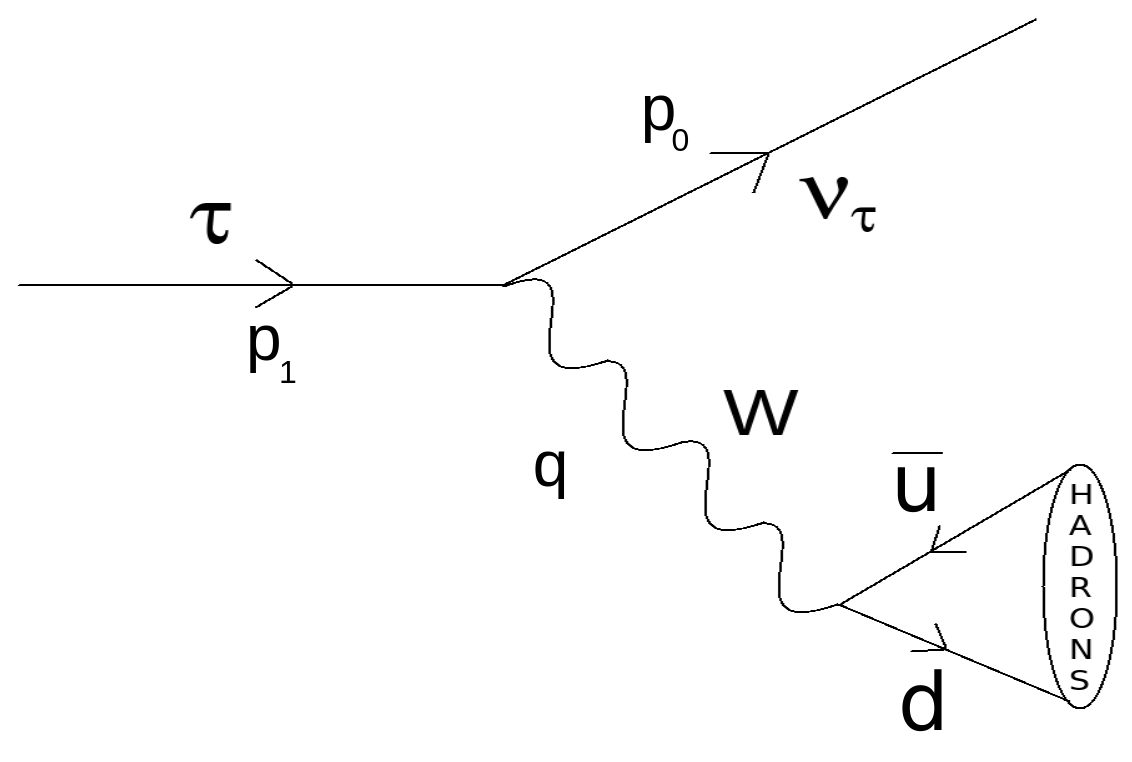}}
\end{center}
\caption{Feynman diagram associated to the hadronic decay of the $\tau$.}
\label{fig:diagram}
\vspace*{-0.5cm}
\end{figure}}
\end{center}

Some of the most powerful methods to study hadronic tau decays involve inclusive observables, which connect experimental spectral functions with imaginary parts of two-point correlation functions (e.g. see \cite{deRafael:1997ea}). A very nice test of asymptotic freedom, which can be translated into a determination of the strong coupling \cite{Braaten:1991qm,Davier:2013sfa,Pich:2013lsa,Boito:2014sta,Pich:2016bdg}, can be performed with the non-strange $V+A$ spectral function. Using also strange data, one can extract information on fundamental parameters such as $m_{s}$ or $V_{us}$ \cite{Gamiz:2002nu,Gamiz:2004ar,Pich:2013lsa,Antonelli:2013usa,Gamiz:2013wn,Hudspith:2017vew}. In this work we use non-strange $V-A$ spectral functions, which, owing to its chiral suppression, become a very nice probe of non-perturbative parameters, such as $\chi$PT couplings or dimensional condensates related with the Spontaneous Chiral Symmetry Breaking of QCD \cite{Donoghue:1993xb,Davier:1998dz,GonzalezAlonso:2008rf,Boito:2015fra,Rodriguez-Sanchez:2016jvw}. 

Another interesting window to test how weak, electromagnetic and strong interactions combine at low energies are non-leptonic kaon decays \cite{Cirigliano:2011ny}, although theoretical uncertainties, due to the complex hadronic dynamics, are typically larger. One of the most interesting and controversial observables in this sector is the CP violating ratio $\varepsilon'/\varepsilon$. While some recent analytical and lattice studies report SM predictions below the experimental measurements \cite{Buras:2015yba,Bai:2015nea}, 
it is well known \cite{Pallante:1999qf,Pallante:2000hk,Pallante:2001he} that the SM prediction agrees with the experimental value, once the pion rescattering in the final state is properly taken into account. This has been confirmed by the recent detailed update of the SM calculation \cite{Gisbert:2017vvj}, which finds a value fully compatible with the experimental one, although with large uncertainties.

The Effective $\Delta S = 1$ Lagrangian in the three-flavour theory is \cite{Buchalla:1995vs}
\be\label{eq:Leff}
\mathcal{L}_{\mathrm{eff}}^{\Delta S=1}\, =\, - \frac{G_F}{\sqrt{2}}\,
 V_{ud}^{\phantom{*}}V^*_{us}\,  \sum_{i=1}^{10}
 C_i(\mu) \, \mathcal{Q}_i (\mu)\, ,
\ee
where the Wilson Coefficients $C_{i}(\mu)$ encode the short-distance dynamics and can be computed with perturbative methods, while nonperturbative hadronic dynamics is captured in the four-quark operators $\mathcal{Q}_{i}(\mu)$. One of the leading contributions to the $\varepsilon'/\varepsilon$ ratio comes from the electroweak penguin matrix elements:
\begin{align}\nonumber
& \langle \mathcal{Q}_{7}\rangle_{\mu}\equiv\langle(\pi\pi)_{I=2}|\mathcal{Q}_{7}|K^{0}\rangle_{\mu}\\&=\langle(\pi\pi)_{I=2}|\bar{s}_{a}\Gamma^{\mu}_{L}d_{a}(\bar{u}_{b}\Gamma^{R}_{\mu}u_{b}-\frac{1}{2}
\bar{d}_{b}\Gamma^{R}_{\mu}d_{b}-\frac{1}{2}\bar{s}_{b}\Gamma^{R}_{\mu}s_{b})|K^{0}\rangle_{\mu} \, , \label{eq:qseven}
\\ \nonumber&\langle \mathcal{Q}_{8}\rangle_{\mu}\equiv\langle(\pi\pi)_{I=2}|\mathcal{Q}_{8}|K^{0}\rangle_{\mu}\\&=\langle(\pi\pi)_{I=2}|\bar{s}_{a}\Gamma^{\mu}_{L}d_{b}(\bar{u}_{b}\Gamma^{R}_{\mu}u_{a}-\frac{1}{2}
\bar{d}_{b}\Gamma^{R}_{\mu}d_{a}-\frac{1}{2}\bar{s}_{b}\Gamma^{R}_{\mu}s_{a}) |K^{0}\rangle_{\mu}\, , \label{eq:qeight}
\end{align}
with $\Gamma^{L(R)}_{\mu}=\gamma_{\mu}(1\mp\gamma_{5})$. Even when there are no known first-principle computations of the different hadronic matrix elements with analytic methods for $N_{C}=3$, one can make use of the fact that the matrix elements of Eq. (\ref{eq:qseven}) and (\ref{eq:qeight}) do not vanish in the chiral limit to connect them to two vacuum condensates by using iteratively the soft-meson theorem. In the chiral limit, {\it i.e.}, at zero momenta, one has \cite{Donoghue:1999ku}:
\begin{align}\label{dg1}
\langle \mathcal{Q}_{7}\rangle_{\mu}&=-\frac{2}{F^{3}}\oo \, ,\\
\langle \mathcal{Q}_{8}\rangle_{\mu}&=-\frac{2}{F^{3}}\left( \frac{1}{2}\oei +\frac{1}{N_{c}}\oo\right) \, .\label{dg2}
\end{align}
with
\begin{align}
\langle \mathcal{O}_{1}\rangle_{\mu}&\equiv\frac{1}{2}\langle 0 |\,\bar{d}\,\Gamma_{\mu}^{L}u\;\bar{u}\Gamma_{R}^{\mu}d \,|0\rangle_{\mu}\, ,
\\ \langle\mathcal{O}_{8}\rangle_{\mu}&\equiv\frac{1}{2}\langle 0 |\,\bar{d}\,\Gamma_{\mu}^{L}\lambda_{i}u\;\bar{u}\Gamma_{R}^{\mu}\lambda_{i}d\, |0\rangle_{\mu} \, ,
\end{align}
where $\lambda_{i}$ are color matrices. 

The functional form of the Operator Product Expansion (OPE) of the $VV-AA$ correlation function \cite{Shifman:1978bx}, $\Pi(s)  \equiv \Pi^{(0+1)}_{ud,LR}(s)\;\equiv\; \Pi^{(0)}_{ud,LR}(s)+\Pi^{(1)}_{ud,LR}(s)$, with
\begin{align}
\label{eq:LRdef}
&\Pi^{\mu\nu}_{ud,LR}(q)\; \equiv\; i \int d^{4}x\, e^{iqx}\;
\bra{0}T\left(L^{\mu}_{ud}(x)R^{\nu \dagger}_{ud}(0)\right)\!\ket{0}
\nonumber\\
&=\; (-g^{\mu\nu}q^{2}+q^{\mu}q^{\nu})\;\Pi^{(1)}_{ud,LR}(q^{2})+q^{\mu}q^{\nu}\;
\Pi^{(0)}_{ud,LR}(q^{2})\, ,
\end{align}
where $L_{ud}^{\mu}(x)\equiv \bar{u}(x)\gamma^{\mu}(1-\gamma_{5})d(x)$ and $R_{ud}^{\mu}(x)\equiv \bar{u}(x)\gamma^{\mu}(1+\gamma_{5})d(x)$, is given at NLO in QCD by:
\begin{align}
\Pi^{(1+0)}(Q^2=-q^2)&=\sum_{p=D/2}\frac{a_{p}(\mu)+b_{p}(\mu)\ln\frac{Q^{2}}{\mu^{2}}}{Q^{2p}} ,\label{eq:nloope}
\end{align}
where $b_{p}$ is $\alpha_{s}$-suppressed with respect to $a_{p}$. The correlator vanishes at all orders in massless perturbative QCD. $a_{1}$ (and $b_{1}$) is suppressed by the light quark masses squared.
The leading contribution of $a_{2}$ is proportional to $\alpha_{s}\hat{m}\langle \bar{q}q\rangle$ and is also numerically negligible. The leading short-distance contribution comes then from operators with dimension $D=6$ \cite{Cirigliano:2001qw}:
\begin{align} \label{unce}
a_3 (\mu) &= 2\left[2 \pi \langle \alpha_s {\cal O}_8 \rangle_\mu +
 A_8 \langle \alpha_s^2 {\cal O}_8 \rangle_\mu +
A_1 \langle \alpha_s^2 {\cal O}_1 \rangle_\mu \ \right] \ ,\nonumber \\
b_3 (\mu) &=2[ B_8 \langle \alpha_s^2 {\cal O}_8 \rangle_\mu +
B_1 \langle \alpha_s^2 {\cal O}_1 \rangle_\mu ]\ \ \, ,
\end{align}
where $A_{i}$ and $B_{i}$ depend on the renormalization prescription and/or on the number of active flavors (they can be found in Ref. \cite{Cirigliano:2001qw}). The $V-A$ correlator is then connected to non-leptonic kaon decays through Eqs. (\ref{dg1}) and (\ref{dg2}) and to inclusive hadronic tau decays, whose associated experimental spectral functions provide its imaginary part. Phenomenological consequences of those relations were studied using mostly tau-decay data in Refs. \cite{Donoghue:1999ku,Cirigliano:2001qw,Cirigliano:2002jy}, where values for those $K\rightarrow \pi\pi$ matrix elements were obtained. Updated data sets \cite{Davier:2013sfa} and further development of techniques to assess the so-called Duality Violation (DV) uncertainties\cite{Chibisov:1996wf,Shifman:2000jv,Cata:2005zj,GonzalezAlonso:2010rn,GonzalezAlonso:2010xf,Boito:2015fra,Dominguez:2016jsq,Rodriguez-Sanchez:2016jvw,Boito:2017cnp} motivate a fresh numerical analysis.

\section{Dispersion relations with polynomial kernel}
The link between the OPE of the correlator, which contains the information on the non-leptonic kaon decay matrix element, and its imaginary part in the region where hadronic tau data allow us to know it is not straightforward, because the former is valid for large Euclidean momenta, $s\equiv -Q^{2}\ll -\Lambda^{2}_{QCD}$ and the latter is only available in the positive real axis at $s<m_{\tau}^{2}$. However, the correlator $\Pi(s)$ is known to be analytic in the whole complex plane except for a cut in the positive real axis. Using that, if we integrate the correlator times an analytic but otherwise arbitrary weight function $\omega(s)$ along the circuit of Figure \ref{fig:circuit}, one finds \cite{GonzalezAlonso:2008rf}
\begin{align}\nonumber
 \int^{s_{0}}_{s_{th}}ds\, \omega(s)\ImPi(s)&-\frac{i}{2}\oint_{|s|=s_{0}}ds\,\omega(s)\Pi(s)\\&= 2\pi f_{\pi}^{2}\omega(m_{\pi}^{2}) \, .\label{eq:poly}
\end{align}
where $s_{th}=4m_{\pi}^{2}$.
\begin{center}
\vspace*{-0.25cm}
{\begin{figure}[t]
\begin{center}
{\includegraphics[width=5.5cm]{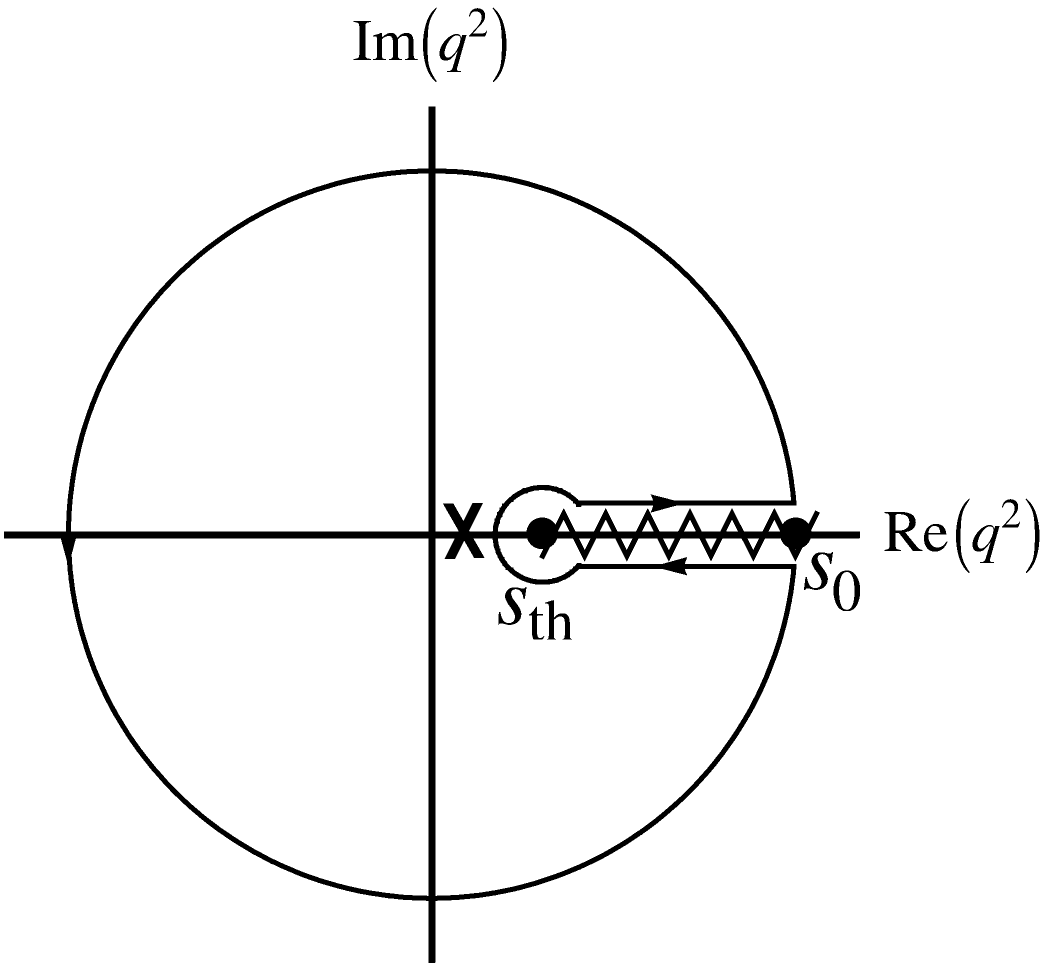}}
\end{center}
\caption{Circuit of integration in Eq. (\ref{eq:poly}).}
\label{fig:circuit}
\vspace*{-0.5cm}
\end{figure}}
\end{center}
In the first term of Eq. (\ref{eq:poly}) one can introduce data, while the second one can be evaluated with the analytic continuation of $\Pi^{\mathrm{OPE}}(s)$. The small differences arising from using the OPE approximant instead of the physical correlator are known as quark-hadron duality violations (DVs) \cite{Chibisov:1996wf,Shifman:2000jv,Cata:2005zj,GonzalezAlonso:2010rn,GonzalezAlonso:2010xf,Boito:2015fra,Dominguez:2016jsq,Rodriguez-Sanchez:2016jvw,Boito:2017cnp}. 

\section{Determination of $\langle (\pi\pi)_{I=2}|Q_{8}| K^{0} \rangle$ in the chiral limit}\label{sec3}
The current size of experimental uncertainties prevents us from working at NLO in $\alpha_{s}$ when extracting the OPE coefficients from tau data (the resolution is not good enough to extract the two $D=6$ terms entering at NLO). As a consequence, we add conservatively (owing to the large value of $A_{8}$), a $25 \%$ of uncertainty to the final result. At that order, a determination of $a_{3}(\mu)$ leads to a determination of $ \langle  {\cal O}_8 \rangle_\mu$. In order to obtain $\langle \mathcal{Q}_{8}\rangle_{\mu}$, one would also need an estimate of $\langle \mathcal{O}_{1}\rangle_{\mu}$. However, this contribution is suppressed by two powers of $1/N_{c}$. This strong suppression has been confirmed by several phenomenological and lattice analyses (e.g. see \cite{Cirigliano:2001qw,Boucaud:2004aa,Blum:2012uk}). Then, one has:
\begin{equation}
\langle (\pi\pi)_{I=2} | {\cal Q}_8 |
K^0\rangle_\mu  =  - {a_{3}(\mu) \over 4\pi\alpha_{s}(\mu) F^{3}}   \, . 
\label{a3vsq8}
\end{equation}
At leading order in $\alpha_{s}$, the determination of $a_{3}$ is equivalent to the determination of $\mathcal{O}_{D=6}$ made in Ref. \cite{Rodriguez-Sanchez:2016jvw}. We have revisited it, introducing some extra tests and trying to implement some small improvements. We proceed as follows:
\begin{itemize}
\item Taking two different ('pinched' at $s=s_0$) weight functions $\omega(s)=\left(1-\frac{s}{s_{0}}\right)^{2}$ and
$\omega(s)= 1-\left(\frac{s}{s_{0}} \right)^2$, we observe good agreement for the obtained values of $a_{3}(\mu)$ for $s_{0}\sim m_{\tau}^{2}$. We also observe a stable plateau for the former. Adding a conservative estimate of DV uncertainties based on the small fluctuations under the change of $s_{0}$, we obtain:
\begin{equation}\label{eq:pinchedres}
a_{3}=(-2.8 \pm 0.9)\cdot 10^{-3} \, \mathrm{GeV^{6}}  \, .
\end{equation}
\item An alternative approach consists in trying to guess how the exact spectral function behaves at $s_{0}>m_{\tau}^{2}$. One pays the price of having to choose a specific parametrization and therefore introducing some model-dependence. We try to relax the model-dependence by allowing data not to obey extrictly the chosen parametrization but imposing they must obey the two Weinberg sum rules (WSRs). The ansatz we use is \cite{Blok:1997hs,Shifman:1998rb,Shifman:2000jv,Cata:2008ru,GonzalezAlonso:2010rn,GonzalezAlonso:2010xf}
\begin{equation}\label{eq.param}
\operatorname{Im}\Pi(s)=\pi\;\kappa \; e^{-\gamma s}\sin{(\beta (s-s_{z}))}\quad s>\hat{s}_{0} \, .
\end{equation}

Following the procedure of Refs. \cite{GonzalezAlonso:2008rf,GonzalezAlonso:2010rn,GonzalezAlonso:2010xf,Rodriguez-Sanchez:2016jvw}
we generate random tuples of parameters $(\kappa,\gamma,\beta,s_{z})$, everyone of them representing a possible spectral function above 
a threshold $\hat{s}_{0}$. If we perform a fit with ALEPH data, we find that there are no significant deviations (p-value above a $5 \%$) 
from this specific model above $\hat{s}_{0}=1.25$ GeV$^{2}$. However, the model is only motivated as an approximation at higher 
energies, where the hadronic multiplicity is higher. As a first constraint, as in Ref. \cite{Rodriguez-Sanchez:2016jvw}, we accept only those tuples that are in 
the $90 \%$ C.L. region ($\chi^{2}<\chi^{2}_{\text{min}}+7.78$). In contrast with Ref. \cite{Rodriguez-Sanchez:2016jvw}, we make a combined fit of the moment used to obtain $a_{3}$ with the WSRs, accepting only those tuples compatible with them (p-value larger than a $5 \%$).\footnote{In this way, correlations between experimental uncertainties when imposing the WSRs and the moment used to extract $a_{3}$ are taken into account.} Our preliminary result is
\begin{equation}\label{eq:modelres}
a_{3}(s_{0})= (-3.5 \pm 1.1) \cdot 10^{-3}\, \mathrm{GeV}^{6}  \, ,
\end{equation}
in good agreement with the result of Ref. \cite{Rodriguez-Sanchez:2016jvw} and with Eq. (\ref{eq:pinchedres}).

\item When assuming a model for the spectral function, as in the previous bullet point, one is changing the assumption of convergence of data to its OPE approximant at $s_{0}\sim m_{\tau}^{2}$, capturing most of the possible DV tails by adding a systematic uncertainty based on fluctuations under the change of $s_{0}$, by the assumption of convergence of data at a lower energy\footnote{This is unfortunately needed in order to fit the free parameters.} to a specific parametrization for the difference between the spectral function and its OPE approximant. A priori, it is unclear to us which procedure should be preferred. One minimal reliability test one should ask to any model, in analogy with the reliability test of independence of the result on $s_{0}$ when directly assuming good convergence of data to its OPE approximant,  is a soft dependence in the choice of threshold $\hat{s}_{0}$. By changing $\hat{s}_{0}$ in the large interval $\hat{s}_{0}\in [1.25,1.9] \, \mathrm{GeV}^{2}$ we have tested that results are stable.
\end{itemize}
 Combining Eqs.  (\ref{eq:pinchedres}) and (\ref{eq:modelres}) and introducing it into Eq. (\ref{a3vsq8}), we find at zero momenta:
\begin{align}
 \langle (\pi\pi)_{I=2} | {\cal Q}_8 |K^0\rangle_{2 \, \mathrm{GeV}}=(1.14\, \pm\, 0.53) \; \mathrm{GeV}^{3} \label{finalchi} \, ,
\end{align}
where the dominant uncertainty originates in $a_{3}$, followed by perturbative errors, estimated as explained above. This value is in good agreement with the results previously obtained by similar approaches \cite{Donoghue:1999ku,Cirigliano:2001qw,Cirigliano:2002jy}. It also agrees with the large-$N_{c}$ estimate\cite{Gisbert:2017vvj}:
\begin{align}
\langle (\pi\pi)_{I=2} | {\cal Q}_8 |K^0\rangle^{N_{c}}_{2 \, \mathrm{GeV}}=2\left(\frac{M_{K}^{2}}{m_{d}+m_{s}}\right)^{2}F_{\pi}\approx 1.2 \, \mathrm{GeV}^{3}
\end{align}
and with the most recent results from lattice simulations \cite{Boucaud:2004aa,Blum:2012uk}.

\section{Using kaon matrix elements from the lattice to improve other tau-based results}
Instead of using inclusive tau-decay data to obtain $K\rightarrow \pi\pi$ matrix elements, one can take advantage of the very precise values for the matrix elements of Eqs. (\ref{eq:qseven}) and (\ref{eq:qeight}) obtained in recent lattice simulations \cite{Blum:2012uk} to determine the coefficients $a_{3}(\mu)$ and $b_{3}(\mu)$. Now we do not have any limitation to work at NLO in $\alpha_{s}$ for the $D=6$ contribution. Using that input and taking $\omega(s)=\left(1-\frac{s}{s_{0}} \right)^{2}$ in Eq. (\ref{eq:poly}), one can obtain a very powerful short-distance constraint for hadronic tau-decay data:
\begin{itemize}
\item Experimental uncertainties, typically dominated by the region near $s_{0}$, are reduced for that weight function.
\item  The first unknown OPE contribution is suppressed both by $8$ powers of the tau mass and by $\alpha_{s}$.
\item Duality Violations are very suppressed for this moment. One would need a very artificial DV shape to make it noticeable. Different model estimates, for example using the tuple corresponding to the minimum in Eq. (\ref{eq.param}), typically predict that they are one order of magnitude below experimental uncertainties at $s_{0}\sim m_{\tau}^{2}$.
\end{itemize}
There are no unknown physical parameters entering into that expression. However, a good way of testing the power of this dispersion relation is simply translating it into a determination of $f_{\pi}$. Even entering in the dispersion relation suppressed by two powers of the tau mass, a quite precise value of this parameter is obtained in Figure \ref{fig:fpi}. As expected, a stable plateau is observed. We find as preliminary result at $s_{0}= m_{\tau}^{2}$:
\begin{align}\nonumber
\sqrt{2}f_{\pi}=(131.6 &\pm 0.9_{\mathrm{exp}} \pm 0.4_{\mathrm{chiral}} \nonumber\ \pm 0.1_{\mathrm{latt}})\, \mathrm{MeV}\\=(131.6 &\pm 1.0)\, \mathrm{MeV} \, ,
\end{align}
where the first uncertainty is experimental, the second stands to the difference between physical matrix elements and the chiral limit values and the last one due to the uncertainty in the lattice input.
\begin{center}

{\begin{figure}[h]
\begin{center}
{\includegraphics[width=8cm]{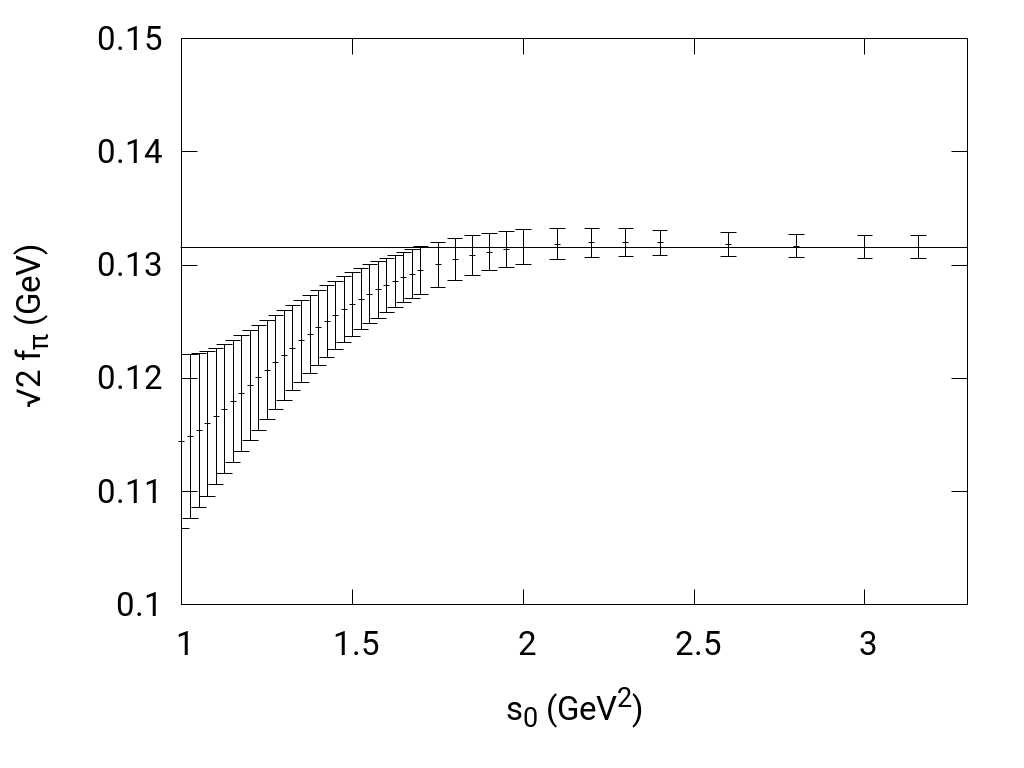}}
\end{center}
\caption{\label{fig:fpi} Equation (\ref{eq:poly}) for $\omega(s)=\left(1-\frac{s}{s_{0}} \right)^{2}$ rescaled so that at $s_{0}$ large enough converges to $f_{\pi}$. An horizontal line with the central value at $s_{0}=m_{\tau}^{2}$ is displayed to guide the eye.}

\end{figure}}
\end{center}

Finally, playing instead the same game as in Section \ref{sec3}, but including the $D=6$ contribution as an external input, we can determine a preliminary value for the $D=8$ condensate:
\begin{equation}
a_{4}=-(0.7 \pm 0.6)\, \mathrm{GeV}^{8} \, .
\end{equation}
\section{Conclusions}
Precise predictions are possible by exploiting relations in the chiral limit between kaon to two-pion matrix elements and vacuum condensates involved in the OPE of certain two-point correlation functions, which can be related to inclusive tau data. From the latter we obtain at zero momenta:
\begin{align}
 \langle (\pi\pi)_{I=2} | {\cal Q}_8 |K^0\rangle_{2 \, \mathrm{GeV}}=(1.14\, \pm\, 0.53) \; \mathrm{GeV}^{3}\, .
\end{align}
Taking instead the $K\rightarrow \pi\pi$ input from the lattice, one can precisely predict some quantities involving inclusive observables dominated by the resonance region around $1 \, \mathrm{GeV}$. For example, they can be translated into a clean determination of $f_{\pi}$ below the per cent level:
\begin{align}
\sqrt{2}f_{\pi}=(131.6 &\pm 1.0)\, \mathrm{MeV} \, ,
\end{align}
or to obtain information about a $D=8$ vacuum condensate
\begin{equation}
a_{4}=-(0.7 \pm 0.6)\, \mathrm{GeV}^{8} \, ,
\end{equation}
in spite of being suppressed by 8 powers of the tau mass.

All the determinations studied here could be improved with future non-strange spectral functions, which in principle could be extracted from Belle-II.

\section*{Acknowledgements}
We  want  to  thank  the  organizers  for  their  effort  to make  this  conference  such  a  successful  event. We are indebted with Vincenzo Cirigliano, Hector Gisbert and Mart\'in Gonz\'alez-Alonso for useful discussion.
We also want to thank Michel Davier, Andreas Hoecker, Bogdan Malaescu, Changzheng Yuan and
Zhiqing Zhang for making publicly available the updated ALEPH spectral functions, with all
the necessary details about error correlations.
This work has been supported in part by the Spanish Government and ERDF funds from
the EU Commission [Grants No. FPA2014-53631-C2-1-P, FPA2017-84445-P and FPU14/02990], by the Spanish
Centro de Excelencia Severo Ochoa Programme [Grant SEV-2014-0398], by the Generalitat Valenciana [PrometeoII/2013/007], by the Swedish Research Council grants contract numbers 2015-04089 and 2016-05996 and by the European Research Council (ERC) under the European Union’s Horizon 2020 research and innovation programme (grant agreement No 668679).
\bibliographystyle{elsarticle-num}
\bibliography{mybibfile}

\end{document}